\documentclass[12pt]{article}
\usepackage{graphicx,amssymb,amsfonts,amsmath,epsfig,color}
\pdfoutput=1
\makeatletter
\DeclareSymbolFont{AMSb}{U}{msb}{m}{n}
\DeclareSymbolFontAlphabet{\mathbb}{AMSb}
\renewcommand{\section}{\@startsection{section}{1}{\z@}%
                                    {-7ex \@plus -1ex \@minus -.2ex}%
                                    {2.5ex \@plus.2ex}%
                                    {\normalfont\large\scshape\centering}}
\renewcommand{\subsection}{\@startsection{subsection}{2}{\z@}%
                                       {-5ex \@plus -1ex \@minus -.2ex}%
                                       {1.5ex \@plus.2ex}%
                                       {\normalfont\normalsize\scshape}}
\renewcommand{\subsubsection}{\@startsection{subsubsection}{3}{\z@}%
                                       {-5ex \@plus -1ex \@minus -.2ex}%
                                       {1.5ex \@plus.2ex}%
                                       {\normalfont\normalsize\scshape}}

\renewcommand\@seccntformat[1]{\ignorespaces\csname #1name\endcsname\space
                               \csname the#1\endcsname.\quad}   % Extra period and name added
\newdimen\captionmargin
\newdimen\captionindent
\newdimen\captionwidth
\newcommand{\captionfont}{\slshape}
\newcommand\@captionlabel[1]{\textsc{#1:}\space}
\long\def\@makecaption#1#2{%
  \vskip\abovecaptionskip
  \captionwidth\hsize
  \advance\captionwidth -2\captionmargin
  \sbox\@tempboxa{\@captionlabel{#1}\captionfont #2}%
  \ifdim \wd\@tempboxa >\captionwidth
    \ifdim\captionindent>\z@
      \advance\captionwidth -\captionindent
      \hskip\captionindent
    \fi
    \hskip\captionmargin
    \parbox[t]{\captionwidth}{\leavevmode\hskip-\captionindent
      \@captionlabel{#1}\captionfont #2}%
  \else
    \global \@minipagefalse
    \hb@xt@\hsize{\hfil\box\@tempboxa\hfil}%
  \fi
  \vskip\belowcaptionskip}
\def\eqnarray{%
   \stepcounter{equation}%
   \def\@currentlabel{\p@equation\theequation}%
   \global\@eqnswtrue
   \m@th
   \global\@eqcnt\z@
   \tabskip\@centering
   \let\\\@eqncr
   $$\everycr{}\halign to\displaywidth\bgroup
       \hskip\@centering$\displaystyle\tabskip\z@skip{##}$\@eqnsel
      &\global\@eqcnt\@ne$\;\hfil{##}$\hfil
      &\global\@eqcnt\tw@$\;\displaystyle{##}$\hfil\tabskip\@centering
      &\global\@eqcnt\thr@@ \hb@xt@\z@\bgroup\hss##\egroup
         \tabskip\z@skip
      \cr}
\ifcase \@ptsize
\or
\or
\fi
  \divide\@tempdima\baselineskip
  \@tempcnta=\@tempdima
\makeatother
\begin{document}

\renewcommand{\theequation}{\arabic{section}.\arabic{equation}}
\renewcommand{\thefigure}{\arabic{figure}}
\newcommand{\gapprox}{%
\mathrel{%
\setbox0=\hbox{$>$}\raise0.6ex\copy0\kern-\wd0\lower0.65ex\hbox{$\sim$}}}
\textwidth 165mm \textheight 220mm \topmargin 0pt \oddsidemargin 2mm
\def\ib{{\bar \imath}}
\def\jb{{\bar \jmath}}

\newcommand{\ft}[2]{{\textstyle\frac{#1}{#2}}}
\newcommand{\be}{\begin{equation}}
\newcommand{\ee}{\end{equation}}
\newcommand{\bea}{\begin{eqnarray}}
\newcommand{\eea}{\end{eqnarray}}
\newcommand{\Identity}{{1\!\rm l}}% Unit Matrix
\newcommand{\cx}{\overset{\circ}{x}_2}
\def\CN{$\mathcal{N}$}
\def\CH{$\mathcal{H}$}
\def\hg{\hat{g}}
\newcommand{\bref}[1]{(\ref{#1})}
\def\espai{\;\;\;\;\;\;}
\def\zespai{\;\;\;\;}
\def\avall{\vspace{0.5cm}}
\newtheorem{theorem}{Theorem}
\newtheorem{acknowledgement}{Acknowledgment}
\newtheorem{algorithm}{Algorithm}
\newtheorem{axiom}{Axiom}
\newtheorem{case}{Case}
\newtheorem{claim}{Claim}
\newtheorem{conclusion}{Conclusion}
\newtheorem{condition}{Condition}
\newtheorem{conjecture}{Conjecture}
\newtheorem{corollary}{Corollary}
\newtheorem{criterion}{Criterion}
\newtheorem{defi}{Definition}
\newtheorem{example}{Example}
\newtheorem{exercise}{Exercise}
\newtheorem{lemma}{Lemma}
\newtheorem{notation}{Notation}
\newtheorem{problem}{Problem}
\newtheorem{prop}{Proposition}
\newtheorem{rem}{{\it Remark}}
\newtheorem{solution}{Solution}
\newtheorem{summary}{Summary}
\numberwithin{equation}{section}
\newenvironment{pf}[1][Proof]{\noindent{\it {#1.}} }{\ \rule{0.5em}{0.5em}}
\newenvironment{ex}[1][Example]{\noindent{\it {#1.}}}

\thispagestyle{empty}

%\begin{flushright}\scshape
%January 2004
%\end{flushright}
%\vskip1cm

\begin{center}

{\LARGE\scshape On the quantum corrected gravitational collapse
\par}
\vskip15mm

\textsc{Ram\'{o}n Torres\footnote{E-mail: ramon.torres-herrera@upc.edu} and Francesc Fayos\footnote{E-mail: f.fayos@upc.edu}}
\par\bigskip
{\em
Department of Applied Physics, UPC, Barcelona, Spain.}\\[.1cm]

\vspace{5mm}

\end{center}

\begin{abstract}
Based on a previously found general class of quantum improved exact solutions composed of non-interacting (\textit{dust}) particles, we model the gravitational collapse of stars.
%Quantum corrections avoid the formation of shell-focussing singularities in the stellar interior. However,
As the modeled star collapses a closed apparent 3-horizon is generated due to the consideration of quantum effects. The effect of the subsequent emission of Hawking radiation related to this horizon
% which continuously evaporates the collapsing object
is taken into consideration. Our computations lead us to argue that a total evaporation could be reached. The inferred global picture of the spacetime corresponding to gravitational collapse is devoid of both event horizons and shell-focusing singularities. As a consequence, there is no \textit{information paradox} and no need of \textit{firewalls}.
%A family of specific regular models of the collapse until reaching the total evaporation is presented.
\end{abstract}

\vskip10mm
\noindent KEYWORDS: Gravitational Collapse, Black Holes, Singularities, Quantum Gravity, Information Paradox.

%%%%%%%%%%%%%%%%%%%%%%%%%%%%%%%%%%%%%%%%%%%%%%%%%%%%%%%%%%%%%%%%
%%%%%%%%%%%%%%%%%%%%%%%%%%%%%%%%%%%%%%%%%%%%%%%%%%%%%%%%%%%%%%%%%%

%%%%%%%%%%%%%%%%%%%%%%%%
%\tableofcontents       %
%\vskip 1cm             %
%%%%%%%%%%%%%%%%%%%%%%%%

\setcounter{equation}{0}

\section{Introduction}

The evolution of a collapsing star has been one of the favorite research topics in theoretical physics for almost a century. The first works in the framework of General Relativity (see, for instance, \cite{O&S}\cite{Vaid1951}\cite{LSM1965}) seemed to show that, under the appropriate circumstances, the formation of a black hole was unavoidable. Moreover, with the later development of the \textit{singularity theorems} \cite{Seno} it was also conjectured that singularities could exist hidden inside black holes.
The contribution of Quantum Theory to this picture was twofold. On the one hand, Hawking showed that the event horizon of a black hole instead of behaving as a one way membrane could emit radiation \cite{Haw75}. This opened the possibility that black holes could evaporate completely. However, this also created a new problem: What happens to all the information that had been swallowed by the Black Hole? If it could not be recovered, then the principle of unitarity of Quantum Mechanics would be violated. This problem is usually known as the \textit{information paradox} (see, for instance, \cite{InfoParadox} and references therein for a more accurate description of the problem).
On the other hand, it is usually expected that quantum theory could avoid the existence of the singularity.
% in a similar way as, in atomic physics, it had avoided the collapse of electrons towards the nucleus of the atom.
%In effect, different approaches to Quantum Gravity
%%(including String Theory, Loop Quantum Gravity, Quantum Einstein Gravity, etc. %%\cite{NonSingularBH})
%have been able to provide solutions for non-singular black holes as well as for some specific %kinds of matter and distributions.
Indeed, some paradigms and heuristic models of \emph{non-singular} gravitational collapse and Black Hole evaporation inspired in different approaches to Quantum Gravity have appeared in the recent literature (see, for example, \cite{Frolov81}\cite{A&B2005}\cite{Hay2006}\cite{Bardeen2014}\cite{Frolov2014}\cite{G&P2014}\cite{H&R2014}\cite{TorresDust}\cite{BCGJ2015}\cite{Haw2014} and references therein). Even some specific, more developed, models are now available (see, for instance, \cite{T&F2013} for the collapse of shells, \cite{BGMS2005} \cite{T&H2011} for homogeneous models or \cite{T&K2015} for scalar field collapse).

Our aim in this work is to analyze the quantum corrected gravitational collapse of matter. We will do it by building and analyzing a family of models that tries to include some of the key physical phenomena involved in the process.
%To reach this goal we will consider quantum corrections in the collapse which will eventually avoid the formation of a %singularity.
In this way, the models will be able to be inhomogeneous and we will take into consideration that the generation of a horizon in the collapsing process requires the emission of Hawking radiation which will tend to evaporate the object. We will integrate the equations describing the collapse in order to analyze the complete stellar evolution. The final global outcome of gravitational collapse will be hypothesized accordingly, as well as the implications for the information paradox and for firewalls.
%and a possible resolution of the information paradox will be provided.

All the computations in this paper are carried out in Planck units where $c=\hbar=G_0=1$.

The work has been divided as follows. In section \ref{SecInt} we describe the general quantum corrected stellar interior. In section \ref{SecExt} we describe a stellar exterior suitable to describe Hawking radiation. The matching of the interior and the exterior solutions and a resolution scheme to obtain the full model can be found in section \ref{SecMatch}. All the details for a  particular model are worked out in section \ref{SecMod}. Finally, the results are discussed in the concluding section \ref{SecCon}, where, based on our results, we conjecture a general picture for the global spacetime of a (non-singular) collapsing star that is able to generate a horizon.

\section{Quantum corrected stellar interior}\label{SecInt}

In order to model the gravitational collapse of the quantum corrected matter we will assume that the whole spacetime will be split into two different regions $\mathcal V =\mathcal V^+ \cup \mathcal V^-$ with a common spherically symmetric time-like boundary $\Sigma= \partial \mathcal V^+ \cap \partial \mathcal V^-$, corresponding to the surface of the star. We choose $\mathcal V^+$ to represent the stellar exterior, while $\mathcal V^-$ will be its interior. In this section we will deal with the exact interior solution.

Recently \cite{TorresDust}, it has been possible to provide the general solution for the quantum corrected collapse of non-interacting particles (dust) in a framework based on the Quantum Einstein Gravity approach to asymptotic safety.
This class of `improved dust interiors' is defined by the metric
\begin{equation}\label{ImprovedLTB}
ds_-^2=-d\tau^2+\frac{R'(\tau,r)^2}{1+2 E(r)} dr^2+ R(\tau,r)^2 d\Omega^2.
\end{equation}
where $\tau$ is the proper time of the particles composing the fluid, $r$ is a parameter that labels every shell of the fluid, $d\Omega^2\equiv d\theta^2+ sen^2\theta d\varphi^2$, the prime indicates partial derivative with respect to $r$ and $R(\tau,r)$, the \textit{areal radius}, is determined by
\begin{equation}\label{Energies}
\frac{\dot{R}^2}{2}=\frac{G M(r)}{R}+ E(r),
\end{equation}
with the overdot indicating partial derivative with respect to $\tau$, $M(r)$ is the \textit{mass distribution}, $E(r)$ is the \textit{energy distribution} and
\begin{equation}
G=G(\tau,r)=\frac{G_0 R^3}{R^3+\tilde\omega G_0 (R+\gamma G_0 M(r))}\label{Gdust}
\end{equation}
is the \textit{running Newton constant}. Here $G_0$ is Newton's universal gravitational constant and $\tilde{\omega}$ and $\gamma$ are constants coming from the non-perturbative renormalization group theory and from an appropriate cutoff identification, respectively.
The preferred value for these constant are $\gamma=9/2$ and $\tilde \omega=167/30\pi$.

By choosing a function $E(r)$ one chooses the total energy per unit mass of the particles in the fluid within a shell of radius $r$. Then,
%(see figure \ref{figBondi})
if $E > 0$ the system would be \emph{unbound}, if $E = 0$ the system would be \emph{marginally bound} and if $E < 0$ the system would be
%\emph{bound}. In fact, the only novelty with respect to the usual LTB solutions is that now the $E<0$ is
\textit{doubly bound}, meaning that there will not only be an upper bound to the particles, but also a non-zero (quantum) lower bound where the shell will bounce.
%
%\begin{figure}
%\includegraphics[scale=1.2]{bondi.eps}
%\caption{\label{figBondi} For a given shell (i.e., a particular $r$) we have plotted a generic %function $V(R)\equiv -G(R) M(r)/R$. From (\ref{Energies}) one deduces that in the $E>0$ case %the behaviour of the shell would be unbounded. It would marginally bounded for $E=0$, since an %expanding shell would reach infinity with $\dot R=0$. Finally, in the $E<0$ case the shell is %\textit{doubly} bounded by a minimum and a maximum radius. Note that, as in the classical %case, there is a limiting lower value for $E$ given by $E_{lim}=-1/2$ and coming from the %signature in (\ref{ImprovedLTB}) and not by the minimum of $V$.}
%\end{figure}
%
%In order to interpret the physical meaning of this solution
%let us suppose that it has been generated by an effective matter fluid in such a way that the %coupled gravity-matter system satisfies Einstein's equations $G_{\mu\nu}=8\pi G_0 T_{\mu\nu}$ %\cite{B&RIS}\cite{BHInt}.

The {\it mass function} \cite{M&S,mass,Hayward}  of this solution is \cite{TorresDust}
\[
\mathcal M^- = M G/G_0
\]
and the apparent 3-horizon (A3H) is the solution of the equation \cite{FST}
\[
R-2 G M=0.
\]
From all the round spheres in the spacetime, only the ones satisfying $R<2 G M$ are closed trapped surfaces.

\begin{figure}
\includegraphics[scale=1]{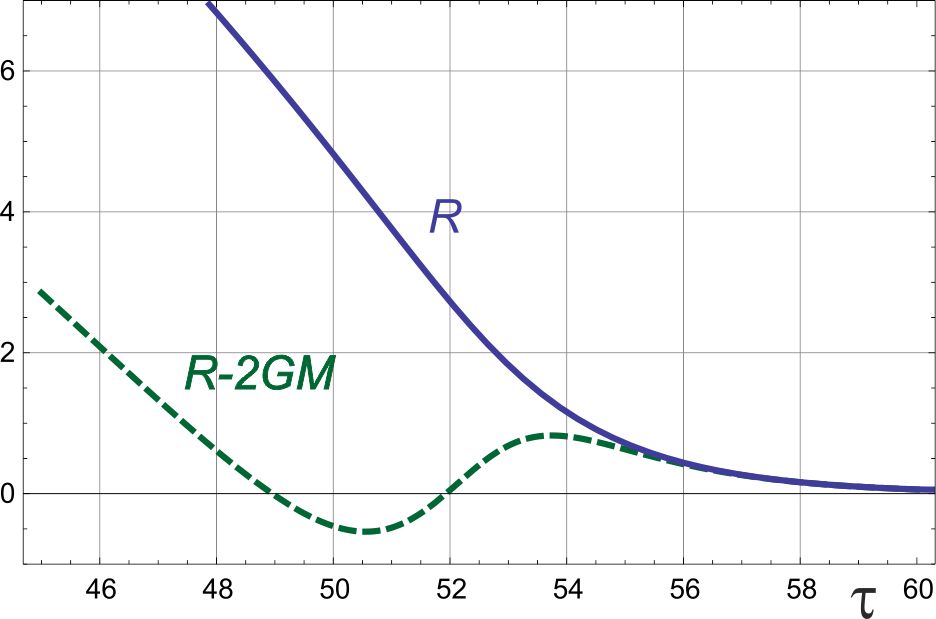}
\caption{\label{figshellevol} In order to illustrate the evolution of a given shell (i.e., a particular $r$) close to the center we have used the particular model that will be introduced in section \ref{SecMod}. Specifically, we show the evolution of the areal radius $R$ of the shell (solid line) as a function of its proper time. Likewise, we show the function $R-2 G M$ along its evolution, what allows to see that starting from positive values of this function the exterior A3H is reached by this shell at $\tau=48.9519$. Then, the shell traverses the region with trapped round spheres until it reaches the interior A3H at $\tau=51.9318$.
%Then the shell strongly decelerates.
}
\end{figure}

The quantum corrected stellar dust behaves essentially as the classical dust far beyond the Planckian regime. In this way, starting from regular initial conditions,
%and in the absence of an A3H,
the model will collapse eventually generating an A3H. However, while in the classical case the later evolution of the shells will force them to fall into a singularity, the quantum corrected shells follow a different path: Once the shells cross the A3H they enter a region of trapped round spheres. While they go through this region the \textit{running} $G$ (\ref{Gdust}) decreases and, as a consequence, weakens the gravitational effects in such a way that, close to the Planckian regime, the shells cross an interior horizon. In this way, the shells enter an inner region of untrapped round spheres. No shell-focusing singularity forms \cite{TorresDust}. We have illustrated this behaviour in figure \ref{figshellevol} by showing the evolution of a particular shell in the particular model that will be used later in section \ref{SecMod}.

Let us now recall the physical content of the solution.
Consider an observer with 4-velocity $\mathbf{u}=\partial/\partial \tau$ and an orthonormal basis
$\{ \mathbf{u},\mathbf{n}, \mbox{\boldmath{$\omega$}}_\theta, \mbox{\boldmath{$\omega$}}_\varphi \}$ such that
\mbox{\boldmath{$\omega$}}$_\theta \equiv R^{-1} \ \partial/\partial\theta$,
\mbox{\boldmath{$\omega$}}$_{\varphi}\equiv (R\sin\theta)^{-1} \ \partial/\partial\varphi$ and $\bf n$ is a space-like 4-vector.
Supposing that it has been generated by an effective matter fluid, the comoving observer with 4-velocity $\mathbf{u}$ will write the effective energy-momentum tensor for the anisotropic fluid as
%vacuum energy-momentum tensor as an anisotropic fluid
%In order to evaluate the physical content of this solution we proceed as with the exterior %solution (sec. \ref{secISS}) obtaining
\begin{eqnarray}\label{TEMint}
\mathbf{T}^- =& \rho\ \mathbf{u} \otimes
\mathbf{u} + p\ \mathbf{n}\otimes \mathbf{n}+ p_{\bot}
(\mbox{\boldmath{$\omega$}}_\theta \otimes \mbox{\boldmath{$\omega$}}_\theta +
\mbox{\boldmath{$\omega$}}_\varphi \otimes \mbox{\boldmath{$\omega$}}_\varphi)\ ,
\end{eqnarray}
where $\rho$ is the effective energy density, $p$ is the effective normal pressure and $p_{\bot }$ is the effective tangential pressure. The field equations provide us with
\begin{equation}
\rho = \frac{(G M)'}{4 \pi G_0 R^2 R'}\ \ , \ \ \
p = -\frac{ M G,_R}{4 \pi G_0 R^2}\ \ \ \mbox{and} \ \ \
p_{\bot } = -\frac{(M G,_{R})'}{8 \pi G_0 R R'}.\label{rho}
\end{equation}
%\begin{eqnarray}
%\rho &=& \frac{(G M)'}{4 \pi G_0 R^2 R'}\label{rho}\\
%p &=& -\frac{ M G,_R}{4 \pi G_0 R^2}\label{pr}\\
%p_{\bot } &=& -\frac{(M G,_{R})'}{8 \pi G_0 R R'}.\label{ptan}
%\end{eqnarray}

\section{Stellar Exterior}\label{SecExt}

With regard to the stellar exterior region $\mathcal V^+$, we will describe it with a portion of a \textit{generalized Vaidya solution} \cite{GVS}.
The spacetime metric for this solution can be written as
\begin{equation}\label{RGISch}
ds_+^2=-\left(1-\frac{2 G_0 \mathcal{M}^+ (u,\bar R)}{\bar R}\right) du^2+2 du d\bar R+ \bar R^2 d\bar \Omega^2.
\end{equation}
where $d\bar \Omega^2\equiv d\bar \theta^2+ \sin^2\bar \theta d\bar \varphi^2$ and $\mathcal{M}^+$ is the mass function of this solution. Depending on the specific chosen mass function this metric has as particular solutions well-known classical solutions as the Schwarzschild solution ($\mathcal M^+=$ constant), the Vaidya radiating solution ($\mathcal{M}^+=\mathcal{M}^+(u)$) and others, but also \textit{quantum improved} solutions such as the improved versions of the Schwarzschild solution (that implies a specific dependence $\mathcal{M}^+=\mathcal{M}^+(\bar R)$) \cite{B&RIS} and the Vaidya solution (that implies a specific dependence $\mathcal{M}^+=\mathcal{M}^+(u,\bar R)$) \cite{B&RIV}\cite{F&T}.

%The apparent 3-horizon for this solution is defined by $\bar R=2 G_0 \mathcal{M}^+(u,\bar R)$.

The characteristics of this solution are very well suited for our model. On the one hand, we have seen that it is able to incorporate the required quantum corrections to the usual classical exterior solutions. On the other hand, this metric is able to describe the specific evolving physical content of the stellar exterior: As the star begins its collapse in a phase previous to the formation of an A3H, we want the stellar exterior to be basically devoid of matter or radiation.
%(so that we could obtain the equivalent to a portion of the classical collapsing dust models).
This can be achieved if one considers for the exterior, at this initial stage, a portion of an improved Schwarzschild solution \cite{B&RIS}. Indeed, as intuitively expected, the model can be built by matching the improved interior and improved Schwarzschild solutions through a stellar surface (/matching hypersurface) $\Sigma$ defined by $r_\Sigma=$constant \cite{TorresDust}. However, once the collapse has reached the stage in which a horizon forms and the stellar surface is located inside it the existence of Hawking radiation in the stellar exterior implies that the exterior solution can no longer be an improved Schwarzschild solution. For example, there will be negative-energy fluxes (composed of radiation and, over time, of massive particles) from the horizon  falling towards $\Sigma$. Therefore, we should consider matching the interior improved solution with a particular portion of the \emph{generalized Vaidya solution} possessing a more general dependence $\mathcal{M}^+=\mathcal{M}^+(u,\bar R)$.

Note that the negative-energy fluxes could also traverse $\Sigma$ and reach interior shells with a depth that depends on the translucency of the stellar interior. We will simplify the model by assuming that the star is perfectly opaque to the radiation so that the negative energy particles will just disintegrate the outermost shells of the collapsing star. In this way, the value of $r$ on the matching hypersurface must decrease with time once the stellar surface has crossed the A3H.

\section{Matching surface with Hawking radiation}\label{SecMatch}

In order to complete our model beyond the point the stellar surface reaches the A3H, let us now match the improved interior solution with the generalized Vaidya solution. $\mathcal V^+$and $\mathcal V^-$ should satisfy Darmois matching conditions on $\Sigma$ \cite{Darmois}\cite{FST}, what implies that the solutions must be such that the first and second fundamental forms of $\Sigma$ must coincide when computed from either $\mathcal V^+$ or $\mathcal V^-$.

%The matching of the interior solution to the generalized Vaidya solution will be performed through a spherically %symmetric time-like hypersurface $\Sigma$.
We choose as a timelike parameter for the matching hypersurface the interior timelike coordinate $\tau$.  The hypersurface will then be described, as seen from the interior, by $r=r(\tau)$.
The normalized 4-vector normal to $\Sigma$ and the stellar surface's 4-velocity can be written on $\Sigma$, from $\mathcal V^-$, as
\begin{eqnarray}
\vec{n}&=&\left(\frac{1+2 E}{R'^2}-\dot r^2 \right)^{-1/2} \left( \dot r\frac{\partial}{\partial\tau}+ \frac{1+2 E}{R'^2} \frac{\partial}{\partial r} \right),\\
\vec v&=&\left(\frac{1+2 E}{1+2 E-\dot r^2 R'^2} \right)^{1/2} \left(  \frac{\partial}{\partial\tau}+ \dot r \frac{\partial}{\partial r} \right).
\end{eqnarray}
As seen from $\mathcal V^+$ ($u=u(\tau)$, $\bar R=\bar R (\tau)$), they take the form, respectively,
\begin{eqnarray}
\vec{n}&=&\left(\left(1- \frac{2 G_0 \mathcal M^+}{\bar R} \right) \dot u^2 -2 \dot u \dot{\bar R} \right)^{-1/2} \left(\dot u \frac{\partial}{\partial u}+ \left[\dot u \left(1- \frac{2 G_0 \mathcal M^+}{\bar R} \right)-\dot{\bar R} \right] \frac{\partial}{\partial \bar R} \right),\\
\vec v&=& \left(\left(1- \frac{2 G_0 \mathcal M^+}{\bar R} \right) \dot u^2 -2 \dot u \dot{\bar R} \right)^{-1/2} \left(\dot u \frac{\partial}{\partial u}+ \dot{\bar R} \frac{\partial}{\partial \bar R} \right).
\end{eqnarray}

Darmois matching conditions and, in particular,
the requirement that the first fundamental forms of $\Sigma$ must coincide
implies that the \textit{areal radii} for the interior ($R$) and exterior regions ($\bar R$) must agree on $\Sigma$ \cite{FST}:
\begin{equation}\label{Rs}
R\stackrel{\Sigma}{=}\bar R.
\end{equation}

Another consequence of the matching conditions are the Israel matching conditions \cite{Israel}\cite{FST} that imply that
the expression $T_{\alpha\beta} n^\alpha (v^\beta-n^\beta)$ should take the same value when evaluated on $\Sigma$ from the interior or the exterior. From the interior, one has
\begin{equation}
T^-_{\alpha\beta} n^\alpha (v^\beta-n^\beta) = \frac{p (1+2 E - \dot{r} R')+\rho \dot{r} R' (\dot{r} R'-1)}{\dot{r}^2 R'^2-(1+2 E)}
\end{equation}
while from the exterior
\begin{equation}
T^+_{\alpha\beta} n^\alpha (v^\beta-n^\beta) = \frac{\mathcal{M}^+,_{\bar R}}{4 \pi \bar R^2}.
\end{equation}
If these two expressions must coincide on the matching hypersurface, then (using (\ref{Rs})) we obtain the \textit{evolution equation} for the stellar surface. For example, in the case of an interior with $E=0$ we have
\begin{equation}\label{evoleq}
\dot{r}\stackrel{\Sigma}{=} \frac{4\pi R^2 p+ \mathcal{M}^+,_{\bar R}}{R' (4 \pi R^2 \rho- \mathcal{M}^+,_{\bar R})}.
\end{equation}
%or, equivalently,
%\begin{equation}
%\dot{r}\stackrel{\Sigma}{=} \frac{2 G,_R M-M,_R G_0}{M,_R G_0 R'-2 G' M-2 G M'}
%\end{equation}

\subsection{Resolution scheme}

In order to model the collapse after $\Sigma$ crosses the A3H we will proceed as follows. First, we will fix the stellar interior or, in other words, we will find a particular solution for (\ref{Energies}). We have to choose particular energy ($E(r)$) and mass ($M(r)$) distributions for the interior together with an initial condition $R(\tau_0,r)$. Then, (\ref{Energies}) will provide us a particular (numerical) solution $R(\tau,r)$. With the help of this solution one can ascertain the particular physical content of the interior (density and pressures) by computing (\ref{rho}). Second, we choose a particular exterior, i.e., a particular mass function $\mathcal M^+$. Finally, the complete set of matching conditions provides us with (\ref{Rs}), the \textit{evolution equation}, the coincidence of \textit{mass functions} at both sides of the matching hypersurface $\Sigma$ \cite{FST}
\begin{equation}\label{masscoin}
\mathcal M^-\stackrel{\Sigma}{=}\mathcal M^+
\end{equation}
and a differential equation for the evolution of the exterior time parameter $u(\tau)$ (see \cite{FST} for details). The complete model can now be obtained from these conditions.
In particular, one will now be able to write $\mathcal M^+,_{\bar R}\stackrel{\Sigma}{=} F(u(\tau),\bar R (\tau,r(\tau)))=F(\tau)$, so that the evolution equation may be numerically integrated. Incidentally, note that in the particular case of the evolution equation (\ref{evoleq}) a \emph{sufficient} condition for the evaporation of the outermost shell of the star ($\dot r<0$) is $F<0$ (since $\rho>0$ and $p<0$).

\section{A model of the last stages of stellar collapse}\label{SecMod}
\begin{figure}
\includegraphics[scale=.9]{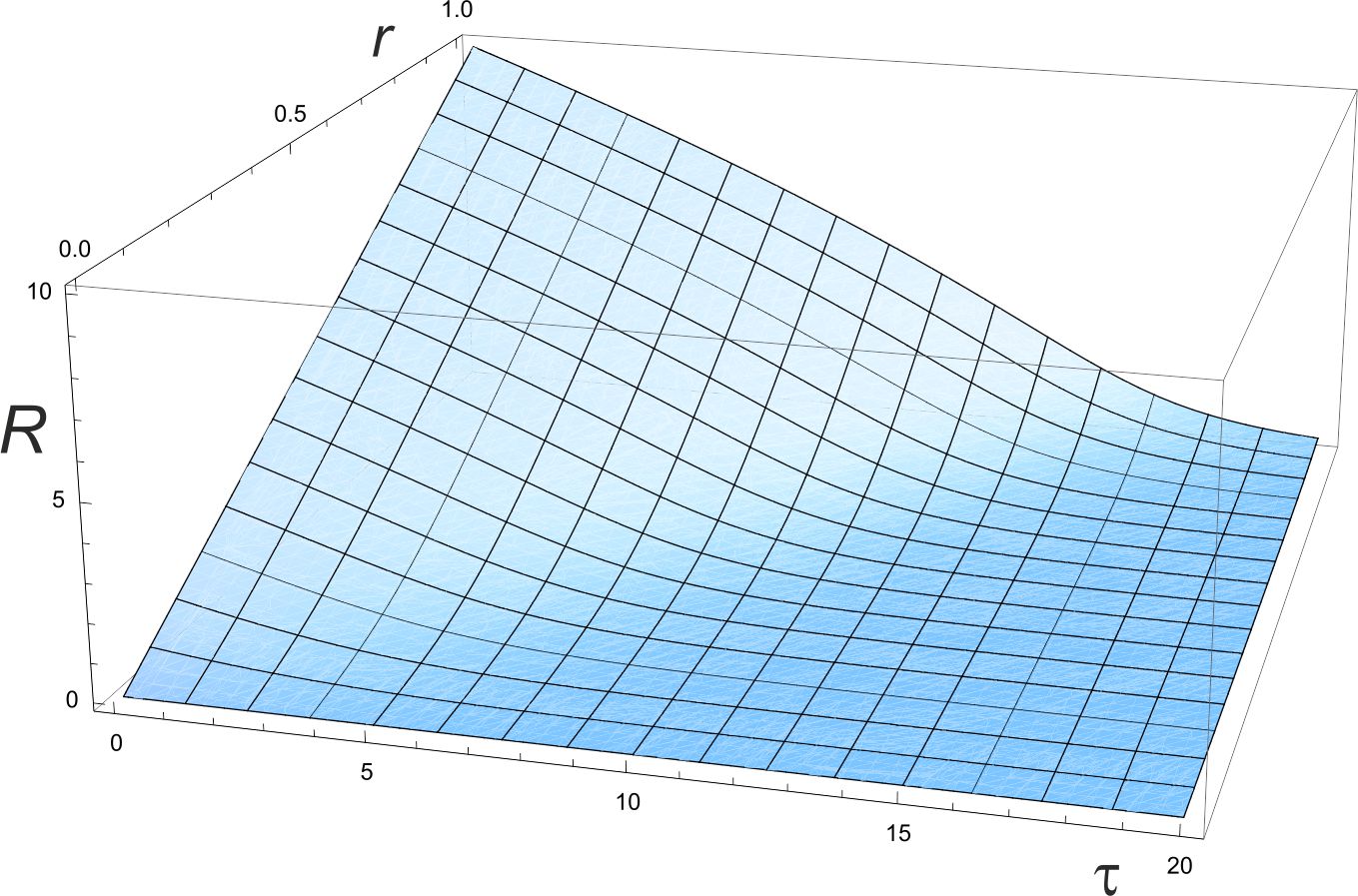}
\caption{\label{figR} A plot of the areal coordinate $R$ for shells around the center of the star, where one can follow the evolution of every shell ($r$=constant) according to its proper time $\tau$. As expected for $E=0$ \cite{TorresDust} the shells collapse towards the center. To plot this figure we have arbitrarily chosen $\varrho=1$ and $\zeta=10$.}
\end{figure}

\begin{figure}
\includegraphics[scale=.8]{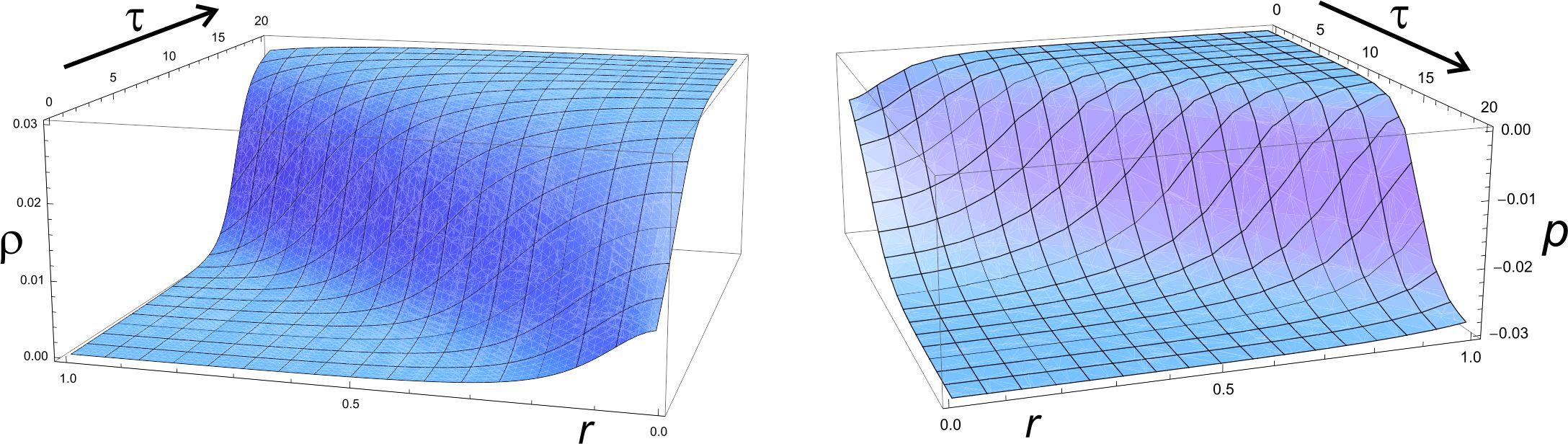}
\caption{\label{figdpp} In this figure we can check how the stellar interior evolves until reaching its maximum value for its density ($\rho_{P}\simeq 0.02994$) and its minimum value for the normal pressure ($p_{P}\simeq -0.02994$). Note that $p$ is always negative due to its quantum origin \cite{TorresDust}.}
\end{figure}

We will choose for the interior a particular simple model in which $E=0$, $M(r)=\varrho r$ and the initial condition is $R(\tau=0,r)= \zeta r$, where $\varrho$ and $\zeta$ are constants. The numerical solution of (\ref{Energies}) for a particular value of the constants provide us with the profile for $R(t,r)$ that can be seen in fig.\ref{figR}.
The implementation of this particular solution in (\ref{rho}) provide us with the effective density and pressures that are shown in fig.\ref{figdpp}. As expected \cite{TorresDust}, note that the modulus of the density and pressures tend towards their maximum finite Planckian value\footnote{Note that the results are of the order of the Planck density and they are in Planck units. In SI units $\rho_{P}\sim 10^{96} kg/m^3$.} given by
\[
\rho_{P} \simeq -p_{P}\simeq -p_{\bot P}\simeq \frac{3}{4 \pi G_0^2 \gamma \tilde\omega } \simeq 0.02994.
\]

%

\begin{figure}
\includegraphics[scale=1]{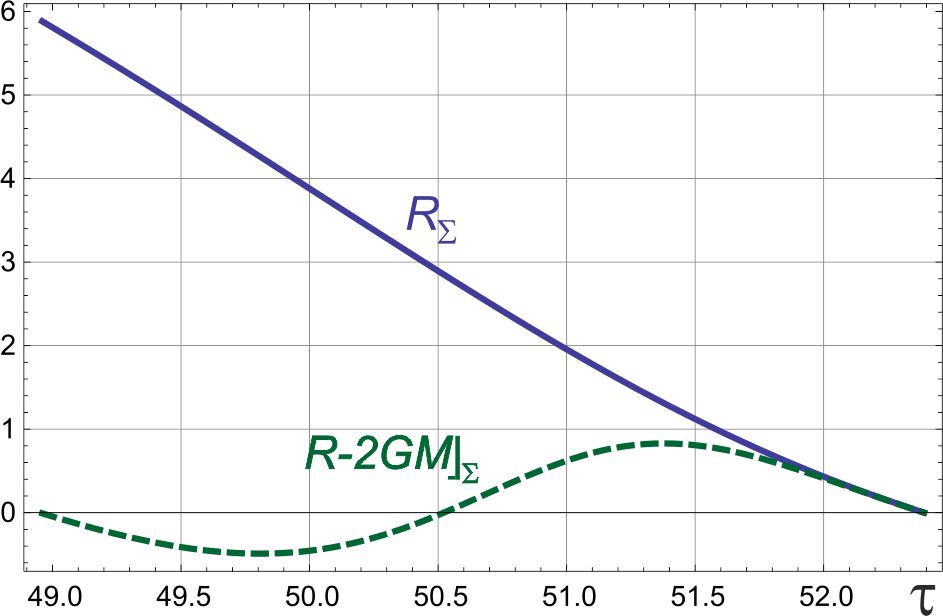}
\caption{\label{figevol} The evolution of the areal radius of the stellar surface $R_\Sigma(\tau)$ (solid line) has been obtained once it crosses the exterior A3H. The value of the function $R-2 G M$ along $\Sigma$ is also shown (dashed line) as a reference. In this particular case, the stellar surface leaves the region of trapped round spheres when it crosses the interior horizon at $\tau=50.5224$. The total evaporation event is at $\tau=52.3912$, when $\Sigma$ reaches $R=0$. The specific initial condition on the A3H is the one in the caption of fig.\ref{figshellevol} and we have arbitrarily chosen the particular value $S_0=0.01$.}
\end{figure}

%As previously stated, before a horizon is created we have chosen an (improved) Schwarzschild solution \cite{B&RIS} for %the exterior. Since previous to the stellar surface crossing a horizon the quantum effects are very small, the surface %will approximately follow the collapsing trajectory from classical theory. Once $\Sigma$ crosses a horizon there will be %negative energy Hawking radiation falling into the star together with backscattered positive energy radiation and we %have chosen to match the interior solution with a particular generalized Vaidya solution.
For our particular example in this section, we choose to work inside the A3H with a particular exterior solution that close to $\Sigma$ takes the form\footnote{Similar and even nicer models can be implemented when the dependence of the exterior mass close to the total collapse is not linear in $R$ but goes as $R^n$ with $n\geq 3$. However, this does not affect significantly our main results.}
\[
\mathcal{M}^+(u,\bar R)\simeq \mu(u)-\frac{\bar R S_0}{2},
\]
where $S_0$ is a constant.
(Incidentally, this solution has been recently used in String Theory \cite{St} in the context of radiating black holes). The solution has the advantage that it is relatively simple to work with for our specific purposes.
In particular, one can work the matching conditions in sequence. For instance, by using the previous numerical results for the stellar interior we can straightforwardly integrate the evolution equation for a given $S_0$
\begin{equation}\label{evoleqpart}
\dot{r}\stackrel{\Sigma}{=} \frac{8\pi R^2 p- S_0}{R' (8 \pi R^2 \rho+ S_0)}.
\end{equation}
If the stellar surface must be a timelike hypersurface then, from (\ref{ImprovedLTB}) it has to satisfy
\[
\frac{-1}{R'}<\frac{dr_\Sigma}{d\tau}<\frac{+1}{R'}.
\]
This requirement is clearly satisfied if one chooses a positive constant $S_0$. Moreover, this choice would also imply the evaporation of the star ($\dot r<0$).
The result of the numerical integration of (\ref{evoleqpart}) is shown in fig.\ref{figevol}. In the same figure we also show the value of the function $R-2 G M$ along the matching hypersurface in order to make explicit how the stellar surface traverses the region of closed trapped surfaces by crossing the inner apparent 3-horizon.
The evolution of the function $\mu$ with the internal time can be now obtained by using (\ref{masscoin}).
%that the \textit{mass functions} at both sides of the matching hypersurface
% $\Sigma$ must coincide $\mathcal M^-\stackrel{\Sigma}{=}\mathcal M^+$\cite{FST}.
This provide us with
\[
\mu(\tau)=\mu(\tau,r(\tau))= M(r(\tau)) G(\tau,r(\tau)) + R(\tau,r(\tau)) S_0/2.
\]
This is a decreasing function with time that reaches zero when the stellar surface completely collapses ($R\rightarrow 0$, what implies $G\rightarrow 0$).

\section{Conclusions}\label{SecCon}

In this paper we have addressed the problem of the quantum corrected gravitational collapse of a star. The modeled stars had an interior composed of non-interacting particles where quantum gravity effects had been taken into consideration (Sec.\ref{SecInt}). These quantum effects make the interior free of shell-focusing singularities. The interior has been matched to an exterior that is suitable to describe quantum corrections outside the star as well as the Hawking radiation that will be produced once an A3H is generated (Sec.\ref{SecExt}). We have taken into account that the negative energy flux of Hawking radiation that will then reach the (opaque) star should evaporate its outermost shells. Therefore, the stellar surface $\Sigma$ that is first described by $r_\Sigma=$constant satisfies, once Hawking radiation reaches it, $\dot r_\Sigma < 0$ (Sec.\ref{SecMatch}). In this way, a total evaporation ($r_\Sigma=0$) due to Hawking radiation seems feasible. In principle, the process would satisfy energy conservation since the initial content of the star that is consumed by the negative energy flux of Hawking radiation is compensated by an equivalent outward flux of positive energy photons and massive particles.

In order to illustrate these ideas we have considered a particular conceptual model (Sec.\ref{SecMod}) and we have analyzed it close to the total evaporation event.
The interior has been chosen so that shell-crossing singularities are avoided in the region under consideration, what is the usual approach since possible shell-crossing singularities are just mathematical artifacts caused by dealing with noninteracting dust shells.
Regrettably, in order to the shells bouncing at the center not to cross between them, this forces us to choose a specific model that quickly collapses\footnote{I.e., shell-crossing is clearly unavoidable in this model if we had allowed the innermost shells to bounce at the center.}. This is not totally satisfactory since one would expect the shells crossing the inner horizon to coexist (an interact) for a longer time while trapped by the inner horizon ($0\leq R <R_{I A3H}$). This \emph{trap} should eventually provide the collapsed star with enough time to evaporate as all the generated negative flux of Hawking radiation reaches it. Therefore, a more complete model would require the description of interacting shells. However, this possibility is not available at the present state of the art and it should be left for future works.

\begin{figure}
\includegraphics[scale=.9]{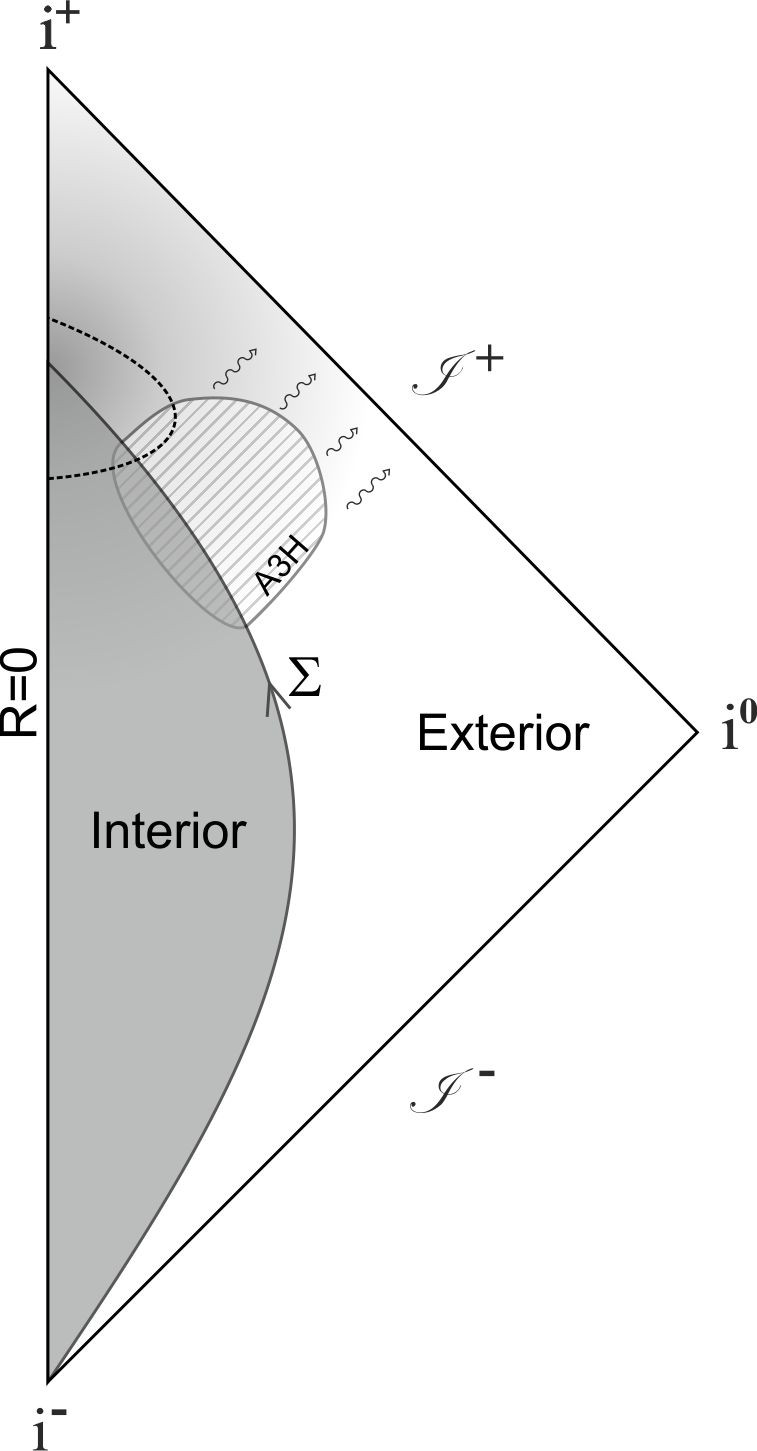}
\caption{\label{figpenrose} The Penrose diagram of gravitational collapse that one could infer from the computations in this work. The stellar interior is limited by the stellar surface (matching hypersurface) $\Sigma$ that collapses towards $R=0$. In the first stages the exterior in basically vacuum and $r_\Sigma=$constant. An apparent 3-horizon (A3H) eventually forms and, in this quantum corrected case, it is closed and contains a region of spherically symmetric closed surfaces with $R>0$ (stripped region) so that one can distinguish between an \textit{interior A3H} and an \textit{exterior A3H}. The formation of the A3H implies the emission of ingoing and outgoing Hawking radiation in the form of photons and massive particles. We have just depicted the outgoing photons by using wavy arrows. The rest of the generated radiation and the remains of the evaporating star are symbolized by a fading grey in the exterior region. The dashed line around the complete evaporation event represents the boundary of the planckian sized region which should be better described by a full theory of Quantum Gravity.}
\end{figure}

The global picture of gravitational collapse that one could extract from our general and particular models, as well as our arguments above, can be seen in figure \ref{figpenrose}.
%
%An isolated star
%%\footnote{In this paper we do not consider the effect of later accretion of matter or radiation.}
%collapses from initial regular conditions. The interior is composed of dust particles and it has basically a vacuum %exterior. Over the course of its (at this stage, practically classical) collapse it generates an apparent 3-horizon. %Then, Hawking radiation is activated so that positive energy particles are emitted from the horizon and negative energy %ones travel towards the stellar surface.
%%As usual, one expects that the radiation is composed mainly of photons in the first stages.
%%However, the evaporation
%In our model the star is opaque to the radiation in such a way that the negative energy particles just evaporate the %exterior shells of the star. As a consequence, the star loses mass while the stellar surface continues its collapse (but %now with $\dot r_\Sigma <0$). Eventually the star completely evaporates and no remnant is produced.
%
It is important to emphasize that the global structure of the spacetime is devoid of an event horizon. Therefore, according to the standard definition there is no black hole.
%One could use, instead, the expression \textit{regular evaporating black hole}.
However, from the point of view of some external observers the object that forms is not distinguishable from a standard evaporating black hole during the evaporation time (about $10^{70}$ years for a solar mass black hole). The collapse is completely regular since no singularity appears. The information loss problem does not have any meaning since there is not a singularity destroying the information, the apparent 3-horizon (that replaces the event horizon) can be traversed by both ingoing and outgoing particles and the spacetime global structure allows the information about the collapsed matter to scape towards null infinity at late times.
%(Among others, even S. Hawking has conjectured \cite{Haw2014} that this behaviour would be the %most reasonable solution to the information loss paradox).
Note that there is no need for a \textit{firewall} in the vicinity of the A3H in order to preserve unitarity. In this way, we have obtained a picture for gravitational collapse that seems reasonable and satisfactory.

Only a warning must be considered: It has been argued \cite{B&RIS}\cite{TorresDust} that the quantum corrections used in this work are accurate for $R$ of the order of Planck's length. Therefore, despite we have depicted the full evolution of the star until its complete evaporation, we cannot guarantee that the last stages of the collapse will be exactly as depicted. The Planckian region where only a full theory of Quantum Gravity could make a totally reliable prediction of the events has been emphasized in fig.\ref{figpenrose}.

%It has been recently conjectured, as a solution of the information paradox, that inside the Planckian region there could %be a bounce generating a \textit{white hole} causally connected with the Universe. However, this requires quantum %gravity effects to be significant even in some regions outside the A3H. This is not easy to justify and there is no %consensus between \cite{H&R2014}\cite{BCGJ2015}

An interesting feature of our model for the global spacetime is that there are external observers in the causal future of the Quantum Gravity domain (fig.\ref{figpenrose}). Not only will these observers be able to recover the apparently lost correlations of matter and radiation that had fallen into the A3H, but they will also be able to collect experimental data from the Planckian region. In this way, in principle, it could be possible to test different theories of Quantum Gravity by observing evaporating Black Holes.

\section*{Acknowledgements}
We acknowledge the financial support of the Ministerio de Econom\'{\i}a y Competitividad (Spain),
projects MTM2014-54855-P

\end{document}